\documentclass[showpacs,amsmath,amssymb]{article}
\usepackage{graphicx}

\begin{document}
\def\kaishu{\CJKtilde \CJKfamily{kai}}
\def\heiti{\CJKtilde \CJKfamily{hei}}
\def\songti{\CJKtilde \CJKfamily{song}}
\def\fangsong{\CJKtilde \CJKfamily{fang}}

\baselineskip 15pt


\title{\bf Theoretical analysis and multi-agent simulation of the ecosystem in Tibet}

\author{{YunFeng Chang$^{1}$\footnote{Email: yunfeng.chang@gmail.com}, BoJin Zheng$^{2}$,Long Guo$^{1}$, Xu Cai$^{1}$}\\
{\small $^{1}$Complexity Science Center,Institute of Particle Physics}\\
{\small Huazhong(Central China) Normal University, Wuhan, 430079,
China.}\\
{\small $^{2}$College of Computer Science}\\
{\small South-Central University For Nationalities, Wuhan,
430074}} \maketitle

\begin{abstract}
Bird Funeral is a strange funeral custom in Tibet of China. The
lamaists hope they can save small fauna such as pikas by
sacrificing themselves to the eagles. But can they save the fauna
by their sacrifice? By theoretical analysis and multi-agent
simulation, we give a negative conclusion that the sacrifice are,
in fact, reducing the population of the pikas. In contemporary
Tibet, the eagle population is reducing drastically, and the
pastures are degenerating. People blame this on the excessive
population of the plateau pikas. We propose a model to explain
this phenomenon. The eagles are dying off because of overgrazing
but not the pikas. We also point out that killing the pikas is
probably unhelpful to recover the pastures but worsen the
degeneration of the pastures. And if people want to recover the
pastures, they have to increase both the population of the
predator and the supply of grass simultaneously.
\end{abstract}

\bigskip

\section{Introduction}
\subsection{A Buddhism Story}
Sivika is the king of Devapati City. Sakra is the Indra of the
devas, the sky-god, the god of the nature-gods, ruler of the
thirty tree heavens. Sakra and another god Visvakarman wanted to
test the Attic faith of Sivika. So they transfigured into an eagle
and a pigeon. The eagle chased the pigeon into the palace. The
pigeon dodged into the king's clothing. So the eagle said to the
king: I am hungry, I have to eat the pigeon and the pigeon can not
escape from me, so please your majesty return it to me. The king
said: I will save all emotional life, the exhausted and fearful
bird dodged in my clothing, I can＊t give it to you. The eagle
replied: If I have no food, I will die; If you want to save all
emotional life, you must save me from starvation. Then the king
cut his leg flesh to the eagle. The eagle said: The meat must have
the same weight as the pigeon. So the king put the flesh and the
pigeon on to a balance, however, the meat was less. The king has
to cut more and more, until all meat on legs, arms out, still
less, at last the king jumped onto the balance. And Sakra was
moved, the earth shook, flower pieces fell from the heaven.

Because of the story above and some other similar Buddhism
stories, the Tibetan think that if they sacrifice his/her body to
the eagle after their death, they can have a better next-life
since it is helpful to save the small fauna. This is the
origination of the bird Funeral in Tibet. But are they doing the
right thing?

\subsection{Ecosystem and Lotka-Volterra Model}

Ecosystem is a complex system, no one can tell exactly the role of
a certain kind of animal in the system. What's more, the same
animal may play different roles under different environments. Here
in this paper we will focus on the situation in Tibet.

The environment in Tibet is very different since the Tibetan
plateau is largely a treeless environment. The open meadows that
constitute the majority part of the plateau make the eco-situation
there very different from other parts of the world. The plateau
pikas there have been considered as\cite{s1,s2} a keystone species
for biodiversity on the plateau, based on a review of the natural
history and ecology of the pika and those species living in
sympatry with it.

But in recent years, the pikas have been blamed for decades as
pest for reducing the forage available for domestic livestock and
for degrading habitant\cite{s3,s4,s5,s6}. And together with the
increasing degradation of the grassland over more than 40
years\cite{s7}, the pikas have become the object of excessive
control effort. So in order to control the population of pikas to
recover the grassland, people began to poison pikas since 1958 and
had escalated greatly by 1962\cite{s6}. Is it a right thing to
recover the grassland by poisoning the pikas since they play an
important role in the ecosystem of Tibet?

 The Lotka-Volterra model was first proposed by Alfred Lotka\cite{s8} in
 1920 to describe the population dynamics of two interacting species, a predator
and its prey. Then in 1926, the very equations was derived by Vito
Volterra\cite{s9} to describe a hypothetical chemical reaction in
which the chemical concentrations oscillate. At the beginning
Lotka-Volterra model was proposed as ecological model to describe
the closed double-species predator-prey interactions that can be
depicted as:
\begin{equation}
\left\{\begin{array}{clcc}&\hspace{-0.3cm}dP/dt=(a-bQ)P\\[0.2cm]
&\hspace{-0.3cm}dQ/dt=(-c+dP)Q
\end{array} \hspace{0.2cm} P,Q\geq0
\right.
\end{equation}
$P$ is the predator population and $Q$ is the prey population in a
closed ecosystem. $a$ represents the natural growth rate of the
prey in the absence of predator, $b$ represents the effect of
predation on the prey, $c$ represents the natural death rate of
the predator in the absence of prey, $d$ represents the efficiency
and propagation rate of the predator in the presence of prey. The
coupling of $PQ$ indicates the interaction between the two
species. Such mathematical models have long proven useful in
describing how populations vary over time and have been extended
to more than two kinds of species.

\subsection{Outline}

We give three Lotka-Volterra models in this paper to answer the
questions above. In the first model, we explain the phenomena of
bird funeral and find that people are doing things with a kind
heart but get the opposite result. It means that the bird funeral
may leads to the reduction of the population of the pikas. Then
why does the population of pikas keep in a high level and the
population of eagles reduced much than before? We solve this
question in the second model. And in the third model, we point out
that killing the pikas is unhelpful to recover the grassland but
probably worsen the degeneration of the pasture. And if we want to
recover the pasture, we must increase the predator population and
the supply of grass simultaneously.

\section{Can bird funeral save the pikas?}

\subsection{Theoretical Lotka-Volterra model analysis}

Assume that the system is closed, the food web can be depicted as:

\begin{figure}[th]
\centering
\includegraphics[width=3.5cm]{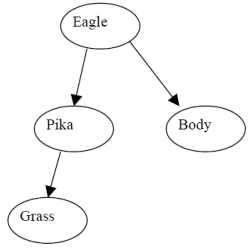} \vspace*{0pt}
\caption{the predator-prey relationship}
\end{figure}

The Lotka-Volterra equation can be obtained as:
\begin{equation}
\left\{\begin{array}{clcc}&\hspace{-0.3cm}dE/dt=(-a_{1}+b_{1}R+f_{1}(B,t))E\\[0.2cm]
&\hspace{-0.3cm}dR/dt=(a_{2}+b_{2}G-c_{2}E)R\\[0.2cm]
&\hspace{-0.3cm}dG/dt=(a_{3}-b_{3}R)G
\end{array}E,R,G\geq0
\right.
\end{equation}
Where, $E$ is the eagle population, $R$ is the pika population, G
is the grass population, $a_{1}$ is the death rate of the eagles,
$b_{1}$ is the rate at which eagles increase by consuming pikas,
$a_{2}$ is the propagation rate of the pikas, $b_{2}$ is the rate
at which pikas increase by consuming grass, $a_{3}$ is the
propagation rate of the grass, $b_{3}$ is the rate at which pikas
destroy grass, $c_{2}$ is the rate at which eagles destroy pikas
and $f_{1}(B,t)$ is the rate at which eagles increase by consuming
bodies at time t.

When analyzing the systems of differential equations, it is often
helpful to consider solutions that do not change with time,
namely, those with $dE/dt=0$, $dR/dt=0$, $dG/dt=0$. Such solutions
are called equilibria, steady-states or fixed points.

Equation (2) would achieve equilibrium when
\begin{equation}
\left\{\begin{array}{clcc}&\hspace{-0.3cm}(-a_{1}+b_{1}R+f_{1}(B,t))E=0\\[0.2cm]
&\hspace{-0.3cm}(a_{2}+b_{2}G-c_{2}E)R=0\\[0.2cm]
&\hspace{-0.3cm}(a_{3}-b_{3}R)G=0
\end{array} E,R,G\geq0
\right.
\end{equation}
Equation (3) gives
\begin{equation}
\left\{\begin{array}{clcc}&\hspace{-0.3cm}R=\frac{a_{1}-f_{1}(B,t)}{b_{1}}\\[0.3cm]
&\hspace{-0.3cm}E=\frac{a_{2}+b_{2}G}{c_{2}}\\[0.3cm]
&\hspace{-0.3cm}R=\frac{a_{3}}{b_{3}}
\end{array} E,R,G\geq0
\right.
\end{equation}
If $\frac{a_{1}-f_{1}(B,t)}{b_{1}}=\frac{a_{3}}{b_{3}}$, then the
equation has a stationary solution. According to the principle of
Volterra, $\overline{R}=\frac{a_{1}-f_{1}(B,t)}{b_{1}}$. If
$\frac{a_{1}-f_{1}(B,t)}{b_{1}}\neq\frac{a_{3}}{b_{3}}$, assume
that $\forall t, E,R,G>0$, the equation is chaotic, we can use the
mean of $R$ to measure the equation. Here
$\overline{R}=\frac{1}{2}(\frac{a_{1}-f_{1}(B,t)}{b_{1}}+\frac{a_{3}}{b_{3}})$.
So if $f_{1}(B,t)\uparrow$, then $\overline{R}\downarrow$. By
coordinate planes analysis, it is very easy to prove that each
coordinate plane is invariant\cite{s10}.

When $\frac{dR}{dt}=0$, we denote $\overline{R}=R_{c}$, so (2)
gives
\begin{equation}
\left\{\begin{array}{clcc}&\hspace{-0.3cm}dE/dt=(-a_{1}+b_{1}R_{c}+f_{1}(B,t))E\\[0.2cm]
&\hspace{-0.3cm}dR/dt=0\\[0.2cm]
&\hspace{-0.3cm}dG/dt=(a_{3}-b_{3}R_{c})G
\end{array} E,R,G\geq0
\right.
\end{equation}
and (5) gives
\begin{equation}
\frac{dE}{dG}=\frac{\frac{dE}{dt}}{\frac{dG}{dt}}=\frac{(-a_{1}+b_{1}R_{c}+f_{1}(B,t))E}{(a_{3}-b_{3}R_{c})G}
\end{equation}
If we assume $f_{1}(B,t)$ as a constant, then equation (6) gives
\begin{equation}
E=KG^{\frac{-a_{1}+b_{1}R_{c}+f_{1}(B,t)}{a_{3}-b_{3}R_{c}}}
\end{equation}
(7) gives, if $f_{1}(B,t)\uparrow$, and if $G$ is invariant, then
$\overline{E}\uparrow$. Because $\overline{R}\downarrow$,
according to $\frac{dG}{dt}=(a_{3}-b_{3}R)G,
\frac{dG}{dt}\uparrow, \overline{G}\uparrow$. That means bird
Funeral is helpful to keep more livestock without changing the
environment much.

From the theoretical analysis above we can get the conclusion that
the bird funeral custom could not save the pikas, but on the
contrary, it makes a lot of pikas die of the increasing eagles.

\subsection{Multi-agent simulation}

We also use multi-agent model to simulate the ecosystem above. In
our model, some simple rules are given: 1. The eagle and the pika
move at random and this movement consumes energy. 2. The eagle and
the pika will die when they don't have enough energy. 3. When the
eagle meets the pika, the eagle will eat the pika and increase the
eagle's energy. 4. The eagle and the pika can reproduce themselves
and their offspring will have 1/2 of the parent energy. 5. When
the pika meet the grass, the pika will eat the grass and increase
the pika's energy. 6. The grass can reproduce at intervals. 7. The
bodies appear at fixed positions at intervals.

We use Netlogo to do the simulation. At the beginning, the number
of bodies is 0, then, after the number of pikas achieves
equilibrium, we set the number of bodies a positive number. We
define the average population of every species as:
\begin{equation}
Apopulation=\frac{\sum_{t=1}^{N}population}{N}
\end{equation}
\begin{figure}[th]
\centering
\includegraphics[width=6cm]{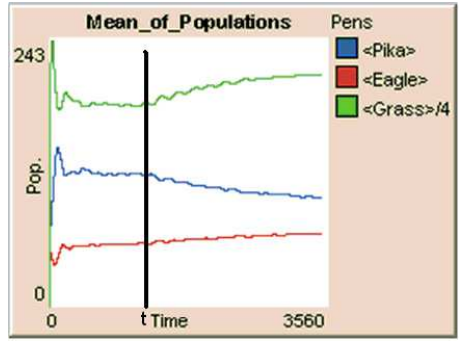} \vspace*{0pt}
\caption{the change of the average population when bodies
increase}
\end{figure}

Fig.2 is the change of the average population when the number of
bodies increase. We can see that the population of pikas are
reducing with the increasing of the number of human bodies. The
reason is that the human bodies will nurture more eagles, and the
increased eagles will eat more pikas. And with the reduction of
pikas, the grass will flourish.

But why are the population of pikas are increasing in Tibet now?
And in fact the population has increased so much that people have
to control it artificially.

\section{Who is the killer of the eagles or the protector of the pikas?}

It is commonly accepted that\cite{s2} the plateau pika of the
QingHai每XiZang(Tibetan) plateau, People＊s Republic of China,
have been considered to be a kind of pest because they have such a
big population that they compete with native livestock for forage
and contributes to pasture degradation\cite{s3,s4,s5,s6}. What is
more, simultaneously, the eagle population is reducing drastically
so that some Tibetan have to abandon the traditional custom of
bird funeral but chooses some other kinds of funeral, e.g.
cremation. Why the population of eagles is decreasing with an
abundant food from pikas? Who is the killer of the eagles and the
protector of the pikas?

\subsection{Theoretical Lotka-Volterra model analysis}

In fact, if we include the influence of livestock, the system will
not work in the simple way as we think. Take the influence of
livestock, the ecosystem can be depicted as:

\begin{figure}[th]
\centering
\includegraphics[width=5cm]{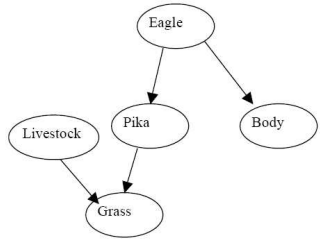} \vspace*{0pt}
\caption{The Predator-Prey Relationship}
\end{figure}

Including the livestock, the differential equation of the
ecosystem in Tibetan can be written as:

\begin{equation}
\left\{\begin{array}{clcc}&\hspace{-0.3cm}dE/dt=(-a_{1}+b_{1}R+f_{1}(B,t))E\\[0.2cm]
&\hspace{-0.3cm}dR/dt=(a_{2}+b_{2}G-c_{2}E)R\\[0.2cm]
&\hspace{-0.3cm}dG/dt=(a_{3}-b_{3}R-f_{3}(L,t))G
\end{array} E,R,G\geq0
\right.
\end{equation}

Where$f_{3}(L,t)$ is the rate at which the livestock destroy the
grass at time $t$.

Equation (9) has ever been discussed in\cite{s10}. If
$\frac{a_{3}-f_{3}(L,t)}{b_{3}}<\frac{a_{1}-f_{1}(B,t)}{b_{1}}$,
the eagle will die out, and if
$\frac{a_{3}-f_{3}(L,t)}{b_{3}}>\frac{a_{1}-f_{1}(B,t)}{b_{1}}$,
the eagle will survive and grow without boundary. And the pika
simply act as a conduit between the top and bottom species.

The real world ecosystem is quite stable relative to human life,
so we can assume that
$\frac{a_{3}-f_{3}(L,t)}{b_{3}}=\frac{a_{1}-f_{1}(B,t)}{b_{1}}$.
Then if $f_{3}(L,t)\uparrow$,
$\frac{a_{3}-f_{3}(L,t)}{b_{3}}<\frac{a_{1}-f_{1}(B,t)}{b_{1}}$,
the population of the eagle will decrease and when
$t\rightarrow\infty$, $E\rightarrow 0$, but the population of the
pikas  won't change much since it is just a conduit.  But once
$E=0$, the system will degrade into the common Lotka-Volterra
Equation, which implies $R\uparrow$.

\subsection{Multi-agent simulation}

We revise the multi-agent model of bird funeral to solve this
problem by adding three new rules: 1. The livestock eat the grass.
2. The livestock move around randomly. 3. The livestock will not
die in the ecosystem without live human.

\begin{figure}[th]
\centering
\includegraphics[width=6cm]{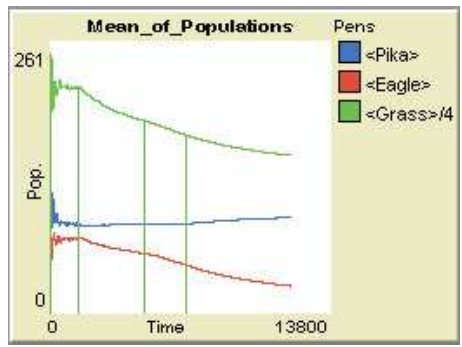} \vspace*{0pt}
\caption{the change of the average populations when overgrazing}
\end{figure}

We can see from Fig.4 that the eagle population will decrease with
the increase of the livestock, so maybe it was the increasing of
the livestock that caused the dying of the eagles. And when the
eagles die out, the population of pikas will increase. However,
according to the actual situation in Tibet, is it the right thing
to recover the grassland by poisoning pika?

\section{Are they doing the right thing?}

Poisoning of pikas began in 1958 and had escalated greatly by
1962\cite{s6}. Till now rodenticide are applied greatly to control
the rodent density. Zhibin Zhang et al.\cite{s11} thought that the
management of livestock is very important for rodent control
because overgrazing is the major factor of causing serious rodent
infestations. In Qinghai, plateau pikas and zokors can be
controlled effectively by rodenticide, followed by the use of
herbicides to control weeds and exclosures to reduce grazing by
livestock, and then re-planting of grass. It is oblivious that the
reduction of grazing by livestock will increase the population of
the eagle, but it won't change the population of the pikas much.
But if the eagles die out, this will increase the population of
the pikas.

Then what is the influence of the poisoning of the pikas?

\subsection{Theoretical Lotka-Volterra Model Analysis}

Take the influence of the poison into consideration, the ecosystem
can be depicted as:
\begin{figure}[th]
\centering
\includegraphics[width=4.5cm]{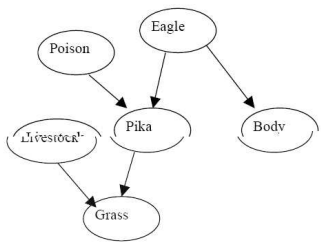} \vspace*{0pt}
\caption{The Predator-Prey Relationship}
\end{figure}

We can obtain the differential equation of Fig.5 as:
\begin{equation}
\left\{\begin{array}{clcc}&\hspace{-0.3cm}dE/dt=(-a_{1}+b_{1}R+f_{1}(B,t))E\\[0.2cm]
&\hspace{-0.3cm}dR/dt=(a_{2}+b_{2}G-c_{2}E-f_{2}(P,t))R\\[0.2cm]
&\hspace{-0.3cm}dG/dt=(a_{3}-b_{3}R-f_{3}(L,t))G
\end{array}   \hspace{0.2cm} E,R,G\geq0
\right.
\end{equation}
$f_{2}(P,t)$ is the rate at which people poison the pikas.
Equation (10) achieves equilibrium when
\begin{equation}
\left\{\begin{array}{clcc}&\hspace{-0.3cm}(-a_{1}+b_{1}R+f_{1}(B,t))E=0\\[0.2cm]
&\hspace{-0.3cm}(a_{2}+b_{2}G-c_{2}E-f_{2}(P,t))R=0\\[0.2cm]
&\hspace{-0.3cm}(a_{3}-b_{3}R-f_{3}(L,t))G=0
\end{array} \hspace{0.2cm}E,R,G\geq0
\right.
\end{equation}
(11) gives
\begin{equation}
\left\{\begin{array}{clcc}&\hspace{-0.3cm}R=\frac{a_{1}-f_{1}(B,t)}{b_{1}}\\[0.2cm]
&\hspace{-0.3cm}E=\frac{a_{2}+b_{2}G-f_{2}(P,t)}{c_{2}}\\[0.2cm]
&\hspace{-0.3cm}R=\frac{a_{3}-f_{3}(L,t)}{b_{3}}
\end{array} \hspace{0.2cm}E,R,G\geq0
\right.
\end{equation}
If $f_{2}(P,t)\uparrow$, $E\downarrow$. And if $E=0$,
$\overline{R}$ will increase drastically. That means poison the
pikas equals poison the eagles, which will cause quicker increase
of the pikas with the decrease of the eagles.

\subsection{Multi-agent Simulation}
We revise the multi-agent model of overgrazing to solve this
problem by adding one new rule: pikas can be killed at intervals.
\begin{figure}[th]
\centering
\includegraphics[width=6cm]{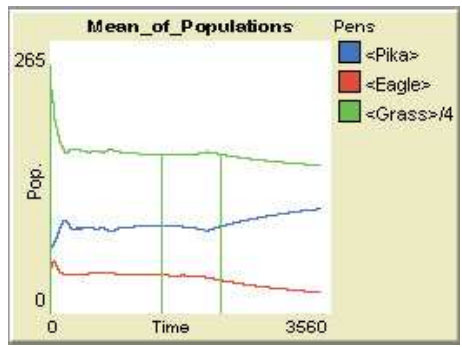} \vspace*{0pt}
\caption{the change of the average populations when poisoning
pika}
\end{figure}

We can conclude from Fig.6 that poison pikas can reduce the
population of the eagle. But after the population of the eagle
decrease to a certain small value, the population of the pika will
increase faster. It means that, if we want to keep the pika in a
low density by using rodenticide, we have to use increasing
rodenticide, otherwise, the pika will boom and compete with the
livestock. This is in accordance with the conclusion of the
theoretical analysis.

\section{conclusion}

In this paper we discuss the ecosystem in Tibet by theoretical
analysis and multi-agent simulation. We begin with the discussion
of the impact of Bird Funeral on the environment of Tibet and get
the conclusion that bird Funeral is helpful to keep more livestock
without changing the environment much though it can not save the
fauna. Then we discuss the environmental impact of overgrazing;
overgrazing can cause the extinction of the eagles, which means
that, to a certain extent, the livestock is the predator of eagle.
Thirdly we discuss the environmental impact of poisoning pikas, we
conclude that this human behavior will extinct the eagle
population.

What is more, from the discussion on the reduction of eagle
species, we can conclude that the ecosystem has two phases, if
$E>0$, the increase of grass will lead to the reduction of pikas,
but if $E=0$ , the pika population will increase. So if we want to
recover the pasture, we must increase the supply of grass on one
hand and make $E>0$ on the other hand. Moreover, we must be
cautious that we should make
$\frac{a_{3}-f_{3}(L,t)}{b_{3}}\geq\frac{a_{1}-f_{1}(B,t)}{b_{1}}$.

We also give out some suggestion according to the actuality in
Tibet for the Tibetan to recover their grassland to protect the
balance of the nature.

Indeed the real-world ecosystems are more complex than the
considered models above. Here we just present some of our toy
models of the ecosystems. We hope we can do some more work on it
in the future.

\end{document}